\documentstyle{article}
\begin{document}

\begin{center}
 {\Large {\bf  Tautomeric Transitions in  DNA }
	 }
	\vspace{2cm} \\
	{\large V.L. Golo and Yu.S. Volkov  \vspace{7mm}\\
	 Department of Mechanics and Mathematics \\
	 Moscow State University \vspace{3mm} \\
         Moscow 119899, Russia
	}
\end{center}

\hspace{1cm}  \vspace{1cm}

\noindent
{\Large {\bf Abstract} } \vspace{5mm}

\noindent
We study the tautomeric transitions in base pairs of DNA
considering elastic properties of DNA as classical and
tunneling of protons as quantum, and show that the dynamics
of the transitions admits of soliton like solutions whose
shape and size strongly depend on the structure of the
double helix.  In particular, we have found that the set of
discrete breathers can be drastically modified by the interplay
of the torsional and elastic constants. Our results may have
a bearing upon substitution mutagenesis within the framework of 
Watson-Crick's approach, and in this respect the breather soliton 
could describe conformations corresponding to point mutations.
The numerical simulation of soliton dynamics suggests that an initial
distribution of base pairs with low probability of mutation per pair
but of a sufficiently large number of base pairs involved,
could move and gather around a site so as to form a set of base pairs
with high probability of mutation, for a period of time
approximately $1  \, \mu sec$. We suggest that the irradiation of DNA at
frequencies of the proton tunneling, that is in infra-red region, could 
cause mutations.

\pagebreak

\noindent
{\bf 1. Introduction}

According to the Watson-Crick hypothesis, \cite{CW1},
the life is built around a symmetrical figure, the double helix of the DNA
molecule (see Fig.1), which comprises
the two strands linked together by purine-pyrimidine base-pairs of
adenine-thymine (AT) and guanine-cytosine (GC),
the four chemicals A,T,G,C existing in various
isomeric forms, or tautomers, that may change into one another (see Fig.2).
Under ordinary conditions the equilibrium shifts towards the amino-form
for adenine and guanine, and the keto-form for thymine and cytosine.
But the imino-form for the adenine and cytosine, and the enol-form for guanine
and thymine are also possible, even though rare;
in fact, they correspond to concentrations of $10^{-4}$  to
$10^{-5}$  moles/liter, \cite{Saenger}.
To see the implications wrought by the tautomeric transitions let us
recall that the sequence of base-pairs constitutes the genetic information
of cell.  It should be noted that adenine will pair only with
thymine by two hydrogen bonds,
and guanine with cytosine with three hydrogen bonds, so that exact copies
of the DNA are produced during the replication (see Fig.3).  But
the complimentarity between the bases is completely changed if a tautomeric
transition takes place; in fact, owing to the structure of hydrogen bonds
other combinations become possible (see Fig.4),
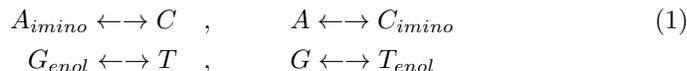
\begin{eqnarray}
 A_{imino} \leftarrow  \rightarrow  C & , & \qquad
 A \leftarrow  \rightarrow  C_{imino}   \\  \label{corr1}
 G_{enol} \leftarrow  \rightarrow   T  & , & \qquad
 G \leftarrow  \rightarrow   T_{enol} \nonumber
\end{eqnarray}
in contrast to the usual and stable ones
$$
      A \leftarrow  \rightarrow   T \qquad , \qquad
      G \leftarrow  \rightarrow   C
$$
Another opportunity for generating  "unnatural"  pairs arises from the
tunneling of protons in hydrogen bonds (see Fig.5), which results
in the formation of the pairs
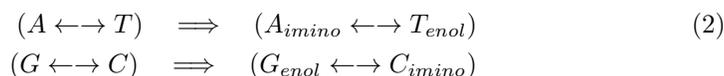
\begin{eqnarray}
 ( A \leftarrow  \rightarrow T ) \quad \Longrightarrow \quad \label{corr2}
 ( A_{imino} \leftarrow  \rightarrow T_{enol} )
 \\  
 ( G \leftarrow  \rightarrow C ) \quad \Longrightarrow \quad \nonumber
 ( G_{enol} \leftarrow  \rightarrow C_{imino} ) \nonumber
\end{eqnarray}
During the replication, tautomeric transition driven by the proton
tunneling in conjunction with the complimentarity according to (\ref{corr1})
may lead to the change of base-pairs
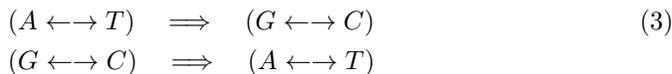
\begin{eqnarray}
 ( A \leftarrow  \rightarrow T ) \quad \Longrightarrow \quad \label{corr3}
 ( G \leftarrow  \rightarrow C )
 \\  
 ( G \leftarrow  \rightarrow C ) \quad \Longrightarrow \quad \nonumber
 ( A \leftarrow  \rightarrow T ) \nonumber
\end{eqnarray}
and result in loss, or corruption, of genetic information, i.e. mutations,
\cite{CW2},  \cite{Loew}.  The specific case given by the diagram
(\ref{corr3}) is called transition mutations; it has the property of
being reversible, i.e. able to go back to the wildlife type.

The arguments given above constitute the main points of the theory of
spontaneous mutations suggested by Crick and Watson,
\cite{CW2}, \cite{Crick}, \cite{topal-fresco}.
It is based on the assumption that the transitory tautomeric shifts of
base-pairs may occur during the replication, i.e.  when two molecules
of DNA are formed from a paired molecule, so that the double-stranded
molecule is split into two single strands, each of which  controls the
synthesis of a new strand complimentary to itself with the help of the
special enzyme called DNA polymerase.  It has been realized that
the latter plays an active role in the selection of bases at replication,
\cite{Auer}, so that it may affect the mutation rates.  Thus, tautomeric
transitions are not a unique cause of mutation;
the situation is more subtle, and many questions,
of quite a classical nature, wait their solutions.  Nonetheless, the
original idea of Watson and Crick still conserves its appeal,
and even more so as its new links with other phenomena related
to the mutagenesis are brought to light (see \cite{Topal}, \cite{Robinson}).
So, Robinson et al, \cite{Robinson} report that the enol tautomer of $iG$,
that is $2^{\prime}-$deoxyisogine, may  form at physiological temperature
($37^o$) and pair with thymine in a Watson-Crick geometry (see Fig.14);
thus, $iG$ present as the nucleoside, results in the formation of
incorrect base-pairs during in vitro replication, \cite{K1},\cite{K2},
\cite{K3},\cite{S1},\cite{S2}. Robinson et al, \cite{Robinson}, suggests that
$iG \cdot T$ pairing may have a bearing on mutagenesis in vivo involving 
tautomers of the common nucleobases. On the other hand,
Fresco et al, \cite{suen}, have found that the imino tautomer $HO^5dCyt$
may serve as an example of an unfavored base tautomer making
for substituting mutagenesis.

Mutations within the framework of the Crick-Watson
model of DNA and in conjunction with the concept of tautomeric transition,
have been drawing attention,  beginning from the early fifties,
\cite{CW1}, \cite{CW2}, \cite{topal-fresco}, to the present time, and
involved the use of condensed matter theory.
So, one of the first papers in this direction was published by
Geracitano and Persico, \cite{Persico}, who suggested that there
should be expected a collective behavior of codons, resembling that
taking place in hydrogen-bonded ferroelectric crystals.

In this paper we intend to look after the interplay between tautomeric
transitions in base-pairs and elastic properties of the double helix.
Since the $\pi-$electrons of the tautomeric rings of
the nucleotides have direct bearing on the interaction of the plates of
adjacent base-pairs,  \cite{Chris}, \cite{HunSand},
we suggest that the tautomeric transition of base-pairs
should substantially influence the distribution of
delocalized electrons of the nucleotides, i.e. the
$\pi-$electrons, and result in deformation of the elastic system of DNA.
It is worth noting that tautomeric transitions may occur
in several base pairs, not necessary adjacent, at a time, and
their dynamics is determined by the proton tunneling.
For one thing the latter is determined  by the electrostatic
interaction, i.e. the dipole forces, between the protons belonging
to adjacent base-pairs, and for another by the elastic system of
the DNA molecule, which should play a role like that of the crystalline
lattice of the polaron theory.
The situation is similar to that considered in the Davydov theory
for the $\alpha-$helix of proteins, \cite{Davy}.
To put these arguments in a more quantitative form
we begin by recalling certain
facts concerning the elastic properties of DNA.  \vspace{1cm}

\noindent
{\bf 2. The torsional modes of DNA}

From the mechanical point of view, the DNA molecule
is a very unusual object.
For one thing one can visualize it as an elastic rod and consider it
within the framework of the mechanics of continuous media,
thus obtaining
reasonable agreement with experiment, \cite{Nelson},
for another it is often necessary to study DNA as a discrete system
similar to crystalline lattice.
In the present paper we have to deal with an
intermediate situation for which a cautious use of the methods of continuous
media is justifiable.  In fact, as far as the transport of
torsional stress (torque) along DNA is concerned, its estimates
obtained by various means diverge widely. The numerical values derived
with the help of the theory of continuous media, \cite{LeviCr}, are of the order
$\tau \propto 10^{-17} dyne \cdot cm $, \cite{Nelson}, \cite{LiuWang},
whereas there is the experimental evidence, \cite{ma}, that it can attain
the value of $\tau \propto 10^{-13} dyne \cdot cm$.
Philip Nelson, \cite{Nelson},
suggested that these deviations could be due to small bends  in the helix
backbone, so that one may assume
$$
   \tau \propto 10^{-17} \div 10^{-13} \quad dyne \cdot cm
$$
As was mentioned above, the interplay between the torsional stress due to
the relative motion of the base-pairs and the proton tunneling is very
important. We shall use the approach worked out in \cite{Peyr} 
and \cite{99:b}
to describe the elastic properties of the double helix.
Thus, the double helix is considered as a one-dimensional lattice of
vectors $\vec y_n$ describing the mutual position of the two strands at sites
corresponding to the base-pair of index n.  It is important that the system
has a twisted ground state characterized by the twist vector $\Omega$,
so that the elastic energy of the molecule can be cast in the form
\begin{equation}
	H_{tor} =
	 \sum_{i=1}^N \: \frac{1}{2} M \left ( \partial_t \vec y_i  \right )^2
	 \quad  +	 \quad
	 \sum_{i=1}^N \: \frac{1}{2} K \left ( \nabla \vec y_i  \right )^2
	 \quad 	 +  \quad
	 \sum_{i=1}^N \: \frac{1}{2} \epsilon \, \vec y_i \,^2 \label{Ten}
\end{equation}
where the first term is the kinetic energy,
the second one the elastic torsional energy
and the last one corresponds to the separation of the two strands.
The covariant derivative that accommodates the torsion of the molecule, reads
$$
  \nabla \vec y_i = \frac{1}{a} \,
		    \left( \vec y_{i+1} - \vec y_i
		    + \vec \Omega \times \vec y_i
					\right )
$$
Here $a$ is the spacing between the adjacent nucleotides,
M is the mass of base-pair. It should be noted that
we are considering a very simplified model and assume that all sites,
corresponding to base-pairs are identical. The subtle question
is the value of the elastic constant $K$; obviously enough it has a direct
bearing on the torque $\tau$ mentioned above, and therefore
its estimate may read
$$
     K \propto 10^{-13} \div 10^{-17} \, erg
$$
It should be noted that the calculations within the framework of molecular
dynamics, (see paper \cite{lavery}  and references therein),
give the upper value for $K$,
i.e. close to $10^{-12} \div 10^{-13} \quad erg$.
For the sake of simplicity, in this paper we shall assume that the
torsion vector $\vec \Omega$ is always
parallel to the axis Oz, that is
$$
	 \vec \Omega = (0,0, \Omega)
$$
and the vectors $\vec y_n$ describe only  transversal motions, that is
$y_n^3 = 0$.

As was mentioned above the tautomeric transitions are driven by the proton
tunneling, and therefore we shall describe them quantum mechanically,
that is the stable amino/keto form corresponding to the
ground state of proton, and the unstable imino/enol one to the excited
state, \cite{Sch1}. In accord with the qualitative character of our
approach we neglect the fact that the tautomeric transitions in question
involve the tunneling of more than one proton, and assign only one
proton to each site of the lattice.

It is important that there are few hydrogen bonds in which the protons
are transferred  towards the imino/keto groups, or if one uses the concept
of the two-level system, the excited states. Therefore, one can consider
the system as being close to equilibrium, or only weakly excited.  This
suggestion is very important for what follows.

We shall describe the states of a base-pair at site $n$ with the Bose
operators $b_n^+,  \, b_n$ that verify the usual conditions
$$
	[ b_n, b_m^+] = b_n b_m^+ \quad - \quad b_m^+ b_n =  \delta_{nm}, \qquad
	[ b_n, b_m] = [ b_n^+, b_m^+] = 0
$$
so that the energy of the protons, neglecting the interaction with the
elastic degrees of freedom, reads, \cite{Sch1}
$$
	H_P= \sum_n E_o b_n^+ b_n \quad + \quad
			 \kappa \sum_n (b_n^+ b_{n+1} + b_{n+1}^+ b_n)
$$
Here $E_o$ is the energy of the tautomeric shift; its estimates depend
on the choice of nucleotide and according to quantum chemistry calculations
vary within the range of $2 \div 10 \, Kcal$,
(see \cite{Saenger} and references therein). The constant
$\kappa$ could be ascribed to dipole interactions between adjacent sites,
similarly to Davydov's theory, \cite{Davy}. Presently, there are no reliable
estimates of its value (see below); by analogy with the Davydov theory
one may assume that it should
correspond to the characteristic frequency of proton , or tautomeric,
excitation of the order $10^{11}$, or less. This figure is generally accepted
(see below).

The central point of the model introduced in \cite{Sch1} is the interaction
between the elastic degrees of freedom of DNA and the tautomeric transitions,
or the proton tunneling in nucleotides;
it reads
$$
	H_I = - \lambda \, \sum_n \left( \nabla \, \vec y_n \cdot \vec h_n \right)
					b_n^+ b_n
$$
An argument  in favor of this choice is that it takes
into account the deformation of positions of adjacent base-pairs and thus
its influence on the $\pi-$electrons of the bases, and therefore,
the tautomeric transitions, or the related excitations of protons.
According to the theory of \cite{HunSand},
the interaction could be appreciable.
Thus, one may suggest that the interaction term could be larger than the
tunneling term in the equation for $H_P$ given above.

Concluding we may state that the total energy within the framework of the
model introduced in \cite{Sch1}  has the form

$$
	H_{total} = H_{tor} \quad + \quad H_P \quad + \quad H_I
$$

\noindent

The conditions discussed above are exactly those used for Davydov's
theory, \cite{Davy}. Let us recall its main points.  It is assumed that
the state of the system can be  described by a trial function that has the
form

\begin{equation}
	 | {\cal D} > = \sum_n \, A_n(t) \cdot b^+_n \, |0 >
	 \label{Dav}
\end{equation}
where $|0>$ is the vector designating the ground state of the system, that is
all the base-pairs, or the protons in the hydrogen bonds, being in the ground
state. The amplitudes $A_n(t)$  are subject to the constraint
\begin{equation}
	\sum_n \, |A_n(t)|^2 \, = 1
	\label{normA}
\end{equation}

At this point it should be noted the vectors $\vec y_n$
describe the dynamics of base-pairs, that is
relatively massive objects, and therefore one may consider them as
classical fields, \cite{Sch1}, \cite{99:b}.
We can derive the size of characteristic frequencies for $\vec y_n$ from expression
(\ref{Ten}) of the elastic energy. In  fact, the mass $M$ is that of the
base-pair, that is of the order $500$ Dalton, and $K$ is of the same order
of magnitude as $\tau$.  Hence, we get the characteristic
velocity $v$ for the $\vec y$ modes
$$
			v \propto \sqrt{ \frac{K}{M} }
$$
Interesting numerical values for the velocity $v$ follow from the equation
indicated above and the rough estimates for $\tau$ or $K$ we have mentioned.
Indeed, for $K \propto 10^{-17} dyn \cdot cm$ or less  we obtain
$$
			v \propto 10^2 \, cm/sec
$$
For wavelengths of a few tens of $\AA$ it gives the characteristic
torsion or phonon frequencies of the order
$$
	\omega_y \propto 10^8 \div 10^9 \: Hz
$$
On the other hand, if we use the values for $K$ provided by
the molecular dynamics simulations, \cite{lavery}, we get the velocity
of excitations of the order $1000 m/sec$, and
$\omega_y \propto 10^{11} \div 10^{12} \, Hz$, as for  ordinary condensed media.
It is instructive to compare the values of $\omega_y$ with the transition
frequencies for tautomeric reactions inside the nucleotides,
$$
\omega_P = \frac{\kappa}{2 \pi \hbar}
$$
The estimates for the latter differ considerably,
\cite{Bodor}, \cite{Leroy}
$$
	 \omega_P \propto 10^6 \div 10^{11} \, Hz
$$
The lowest estimate, $10^6$ Hz appears to be  not unreasonable
(V. Benderskii, and J.L.Leroy, personal communications).

The relative sizes of $\omega_P$ and $\omega_y$ are important
for choosing the right approximation for the model.
In fact, if we are at the lowest end of the spectra $\omega_P$,
then according to the estimate for $\omega_y$
obtained above the characteristic times for the acoustic modes are at
least by an order of magnitude smaller than for the protons. In this case,
we may suggest that the elastic system should follow the motion of the
protons in hydrogen bonds, adjusting itself to it, so that the situation is
similar to that of the Born-Oppenheimer approximation in the atomic theory.

Thus, we assume, as in paper \cite{Sch1}, that the adiabatic
approximation is valid, and therefore we may neglect the kinetic energy
of the elastic system and take into account only its potential energy
generated  by the field $\vec y_n$. Then we are in
a position to apply Davydov's method, \cite{Davy}, that is to
calculate the mean value
$$
		U_{eff} = < {\cal D}| H_{tor} + H_I |{\cal D}>
$$
find the minimum, $\vec y_n^{(o)}$ of $U_{eff}$  with respect to
$\vec y_n$, substitute it into the equation for the total energy
$H_{total}$ so as to get the effective Davydov hamiltonian ${\cal H}_D$,
which depends only on the operator variables $b^+_n, \, b_n$, the classical
variables $\vec y_n$ having disappeared through the minimization. Thus,
we obtain an equation that has the form of the Schr\"odinger one
\begin{equation}
	i \hbar \frac{\partial}{\partial t} \, | {\cal D}>  =
					 {\cal H}_D | {\cal D} > \label{SD}
\end{equation}
and in which the wave function $| {\cal D} >$
should be of the form (\ref{Dav}). 
The assumption that the excited states correspond to the set of two-level
systems is accommodated by the requirement that the operators $b^+_n$
are allowed only in the first power.
It results in a system of equations,
called the Davydov equations, for the amplitudes $A_n$, which one obtains
on equating the coefficients at $b^+_n$ on
both sides of (\ref{SD}), ( see \cite{Davy} for the details ).
All the necessary calculation for the
model of the DNA we employ, had been done in \cite{Sch1}, on which
the present paper relies.

The Davydov hamiltonian for our problem reads

\begin{eqnarray}
  H_D &=& \sum_n E_0 b_n^{+}b_n
       -\sum_n \kappa (b_{n+1}^{+}b_n + b_n^{+}b_{n+1}) \nonumber \\
   & & -\frac{\lambda^2}{k} \sum_n|A_n|^4 
       -\frac{\lambda^2}{k} \sum_n|A_n|^2 b_n^{+} b_n \nonumber \\
   & & +\frac{\lambda^2}{2k} \frac{\epsilon a^2}{k\Omega^2}
     \sum_{m,n}cos^{|m-n|}\phi \cdot cos \left[ (m-n)(\phi-\alpha) \right]
     |A_m|^2 |A_n|^2 \nonumber \\
   & & +\frac{\lambda^2}{2k} \frac{\epsilon a^2}{k\Omega^2}
     \sum_{m,n}cos^{|m-n|}\phi \cdot cos \left[ (m-n)(\phi-\alpha) \right]
     |A_n|^2 b_m^{+} b_m \nonumber
\end{eqnarray}

\noindent
and the equation for the amplitudes $A_n$

\begin{eqnarray}
  i \hbar \frac{\partial}{\partial t} A_n &=&
   E_0 A_n
   -\kappa(A_{n+1} + A_{n-1}) \nonumber \\
   & & -\frac{\lambda^2}{k}|A_n|^2A_n 
       -\frac{\lambda^2}{k}(\sum_{m}|A_m|^4)A_n \nonumber \\
   & & +\frac{\lambda^2}{k} \frac{\epsilon a^2}{k\Omega^2}
     ( \sum_{m_1,m_2}cos^{|m_1-m_2|}\phi 
     \cdot cos \left[ (m_1-m_2)(\phi-\alpha) \right]
     |A_{m_1}|^2 |A_{m_2}|^2 ) A_n \nonumber \\
   & & +\frac{\lambda^2}{k} \frac{\epsilon a^2}{k\Omega^2}
     ( \sum_{m}cos^{|m-n|}\phi 
     \cdot cos \left[ (m-n)(\phi-\alpha) \right]
     |A_{m}|^2 ) A_n \nonumber
\end{eqnarray}

see \cite{Sch1} for the details.  \vspace{1cm}

{\bf 3. The numerical simulation}

We introduce the reduced variables $B_n$ according to the equation

$$
    A_n = e^{-\frac{i}{\hbar} E_0 t} B_n(t)
$$

\noindent
and cast the equation for $A_n$ in the form

\begin{eqnarray}
  i \hbar \frac{\partial}{\partial t} B_n &=&
   -\kappa(B_{n+1} + B_{n-1})
   -\frac{\lambda^2}{k}|B_n|^2B_n \label{Sch2NL} \\
   & & -\frac{\lambda^2}{k}(\sum_{m}|B_m|^4)B_n \nonumber \\
   & & +\frac{\lambda^2}{k} \frac{\epsilon a^2}{k\Omega^2}
     ( \sum_{m_1,m_2}cos^{|m_1-m_2|}\phi |B_{m_1}|^2 |B_{m_2}|^2 )
     B_n \nonumber \\
   & & +\frac{\lambda^2}{k} \frac{\epsilon a^2}{k\Omega^2}
     ( \sum_{m}cos^{|m-n|}\phi |B_{m}|^2 )B_n \nonumber
\end{eqnarray}
Here
$$
  \phi = \arctan \Omega
$$

The non-local terms are a consequence of the structure
of double-helix, often neglected in considering the dynamics of DNA,
\cite{Stari}.

Introduce the characteristic frequencies
\begin{equation}
  \omega_P = \frac{\kappa}{2 \pi \hbar}, \quad
  \omega_T = \frac{\lambda}{2 \pi \hbar}, \quad
  \omega_{tor} = \frac{K}{2 \pi \hbar}
  \label{Freq}
\end{equation}
and the dimensionless time
$$
   \Upsilon = t \cdot \omega_P
$$
It should be noted that the frequencies $\omega_y$ and $\omega_{tor}$
are not identical, $\omega_y \neq\omega_{tor}$.
Then the Davydov equation takes the form
\begin{eqnarray}
  i \frac{\partial}{\partial \Upsilon} B_n &=& 
   -(B_{n+1} + B_{n-1})
   -W|B_n|^2B_n \label{schnl} \\
   & & -W (\sum_{m}|B_m|^4)B_n  \nonumber \\
   & & +W \Lambda
     ( \sum_{m_1,m_2}cos^{|m_1-m_2|}\phi |B_{m_1}|^2 |B_{m_2}|^2 )
     B_n \nonumber \\
   & & +W \Lambda
     ( \sum_{m}cos^{|m-n|}\phi |B_{m}|^2 )B_n \nonumber
\end{eqnarray}
in which
\begin{eqnarray}
  W &=& \frac{\omega_T^2}{\omega_P \cdot \omega_{tor}} \label{W}\\
  \Lambda &=& \frac{\epsilon a^2}{k \cdot \Omega^2}     \label{Lambda}
\end{eqnarray}

Now we aim at making the numerical simulation of equation (\ref{schnl})
for various values of the parameters $W, \, \Lambda$, looking for solutions
of the soliton type.  We use the term soliton in a sense close to that used by
applied scientists, i.e. a solution different from zero in a finite region
of space, whose {\it size we shall call the size of soliton}
and preserving its shape for very long periods of time.
For some values of $W, \, \Lambda$ it has the form identical to the
usual one, i.e. corresponding to the non-linear Schr\"odinger equation,
but generally our solitons are different.  The standard definition
suggests that it be of the form
\begin{equation}
 Y(x,t) = e^{i(qx - \nu t)}\, \psi(x -  vt)  \label{sts}
\end{equation}
in which $\psi$ is a real function.  It is by no means clear that solutions
that we suppose to be solitons, always have the form given by equation
(\ref{sts}).

The parameter $\Lambda$ is a quantitative characteristic that enables us to 
take into account the structure of the double helix, and also the relative 
size of the torsional and deformation energies. In fact, $\Lambda$
determines the magnitude of the nonlocal terms in equation (\ref{schnl}),
and in this respect it is worthwhile to note that for certain values of
$\Lambda$ and W we have not been able to find soliton solutions, e.g.
$\Lambda = 0.2$ and $W = 2$, at least for physically reasonable sizes of
solitons, i.e. less than 100 base pairs.
But it is important that generally the condition
$\Lambda \neq 0$ does not forbid the existence of solitons, and its
influence only results in the size of soliton becoming larger, which
is quite natural, for $\Lambda$ represents non-local terms in equation
(\ref{schnl}). The general case of soliton with $\Lambda$ not equal to zero,
even though small, is illustrated in Fig.8.

To illustrate the general situation let us consider
the two special cases (for the details of calculation see Appendix).

1. Stationary solutions in the sense that the absolute value,
$|B_n(t)|$ does not depend on time. For the usual solitons given by
(\ref{schnl}) this requirement means that the velocity $v = 0$.
The typical case is illustrated in Fig.9, for $W = 10$ and $\Lambda = 0.5$.
The half-width of soliton is equal one spacing between base-pairs,
that is the solution is extremely narrow, and according to our main hypothesis
it must correspond to the tautomeric transition of a base-pair.
The very interesting case is illustrated in Fig.10, $W = 5$ and 
$\Lambda = 0.5$. There is a central peak of half-width $1.5 \cdot a$ which
stands still, and two symmetrical wave packets, moving in opposite 
outward directions. The distance traveled by these wave packets during
0.018 msec is equal to 33 base pairs.

2. The usual solitons given by (\ref{schnl}). The half-width of these solitons 
may be several tens of base-pair spacings, and thus they could correspond 
to tautomeric transitions taking place in adjacent base-pairs.
The typical cases are illustrated in Fig.11, 12.
The distance traveled by soliton in Fig.11 during 0.53 msec is equal to
707 base pairs, and the distance traveled by soliton in Fig.12 during 
0.577 msec is equal to 491 base pairs. It is interesting to note that 
these solitons move, even though slowly.
Hence, one might suggest the picture of  tautomeric
transitions moving along the DNA-molecule.

Both types of solutions indicated above are stable
with respect to perturbation of W and $\Lambda$.

Perhaps, the most characteristic feature of discrete non-linear
Schr\"odinger equation is solutions that periodically oscillate in time and
decay exponentially in space, or breathers, \cite{flach}.
From a purely qualitative point of view the existence of
breathers can be inferred from a truncated version of equation  (\ref{schnl}).
Let us neglect all the terms on its RHS except the first two,
that is consider

$$
  i \hbar \frac{\partial B_n}{\partial t} =
  - (B_{n+1} + B_{n-1})  - W |B_n|^2 B_n
$$

\noindent
and look for $B_n$ such that

$$
   B_n = e^{i \nu t} a_n
$$

\noindent
$a_n$ being real.  Next, cast the equation for $a_n$ in the form

$$
   - (a_{n+1} - 2 a_n + a_{n-1})  -  [W a_n^3  + (2 - \epsilon)] a_n = 0
$$

\noindent
Suppose that the soliton we are looking for is large enough so that
we may change the expression  $a_{n+1} - 2 a_n + a_{n-1}$ for the second
derivative.  Thus we obtain the equation

$$
  a^{\prime \prime} + [W a^3  + (2 - \epsilon)] a  = 0
$$

\noindent
or the conservation law for one dimensional motion with the
effective potential

$$
   V = \frac{2 - \epsilon}{2} a^2  \, + \, \frac{W}{4}  a^4
$$

\noindent
The soliton solution exists for $\epsilon \geq 2$, and its size
tends to infinity as $\epsilon \rightarrow 2$.  On the other hand
for large $W$ we may expect thin solitons.

The key point is that the nonlocal terms generated by the double helix
bring serious modifications to the picture given above.

We may infer from the examples given above that the dimensionless constants
$W$  and $\Lambda$ play a crucial role in determining the form of solitons for
equation  (\ref{schnl}).  The general situation to the effect is illustrated
in Figs.16 (a)-(b), in which the horizontal axis corresponds to values of $\nu$,
that is the soliton frequency measured in units of $\omega_P$.
It should be noted that $\nu$ is well defined for solitons of the form
given by equation (\ref{sts});  in contrast, there is a fine structure
in the frequency spectrum of breathers, \cite{flach}, so that $\nu$
turns out to be only a rough characteristic.  In fact, in Figs.16 (a)-(b) the
values of $\nu$ are determined  to within about one hundredth of $\omega_P$.
As is shown in Figs.16 (a)-(c) for $\Lambda = 0, \, 0.1, \, 0.15$, respectfully,
the set of $W$ and $\nu$, for which there are solitons or breathers,
consists of a line corresponding to breathers, and a region, or domain,
for moving solitons.  The line serves also as a right border for
the region of moving solitons.  The lower border of the soliton region
is not strictly defined owing to the fact that there are solitons
for values of $W$ and $\nu$ lower than the borders but of
sizes greater than $100$ base pairs, that is outside the physical context
of our problem. The upper left part of the border is determined by solitons
turning out to be unstable for values of $W$ and $\nu$ beyond the boundary. 
We see that the soliton region is decreasing as $\Lambda$
grows, and for  $\Lambda = 0.2$, Fig.16 (d), there are only breathers, at least
under the constraint of their size being less than $100$ base pairs.
It is worth noting that the equation (\ref{schnl}) derived in \cite{Sch1}
is valid only for small $\Lambda$.

We would like to draw attention to a class of solutions that are not solitons,
but nonetheless may have a bearing upon the dynamics of tautomeric 
transitions. A solution of the type is illustrated in fig.13. It is 
characterized  by an initial set of amplitudes $B_n(t)$  which is a broad 
distribution of the size of 80 base-pair spacings; after the period of time
0.017 msec, it focuses itself on a narrow peak of half-width of one spacing.
The peak exists for the brief period of time 0.002 msec, and next breaks 
down into a broad distribution again, i.e. a kind of partial self focusing
is taking place. Thus, there may exist low probability
tautomeric transitions distributed over wide areas of the molecule, and
which may collapse into a small region of the molecule, and stay there 
for a period of time, brief but perhaps sufficient to cause mutation.

Finally, we wish to tell that our simulations used the standard numerical
methods, i.e. the trapezoid, the Adams-Boshoft and the Adams-Moulton
of the fourth order, the algorithm for stiff systems.  An important test
has been the conservation of the normalization condition
\begin{equation}
  \sum_n |B_n(t)|^2 = 1 \nonumber
\end{equation}
For testing the precision of our algorithms we have also used calculations
backwards in time.

\vspace{1cm}

\noindent
{\bf 4. Conclusions}

As was shown above the dynamics of tautomeric transitions in DNA depend 
on elastic properties of the latter and proton tunneling in base pairs;
$W$ and $\Lambda$ serving as indicators for possible regimes.
Our numerical simulation suggests that the interesting
tautomeric dynamics may happen for $W \ge 1$. This allows for
sufficiently wide range of material constants of DNA so as to hope the
phenomenon's taking place. The second constant, $\Lambda$, provides
a quantitative characteristic for the part played by the double helix;
it can totally modify the structure of solitons
corresponding to tautomeric transitions.

Depending on the value of W one may expect the existence of two quite
different kinds of soliton dynamics. The first one belongs to solitons
that move at velocity of several $0.01 \, cm/sec$, have a size of
several tens of base-pairs,  and the second of stationary solutions,
or breathers, that have a form of peaks over a few base-pairs.
We may suggest that the second type of solutions
correspond to  point mutations, whereas the first one
may describe tautomeric transition moving along the chain of
double helix. According to the Crick-Watson approach,
there may happen mutations related to the transition.
Thus, one may suggest that an action imposed on a set of nucleotide
in a region of the molecule might generate mutations in a different region
owing to the motion of excitations corresponding to proton tunneling.

It is alleged to be known that by substituting the "artificial"
nucleotides instead of the natural ones, e.g. brom-uracil for thymine
(see Fig.6,7), one can increase dramatically the rate of mutations; 
this could be due to the increase of tautomeric transitions inside  
base-pairs. At any rate, it is worthwhile to study the interplay between 
the rate of such transitions and mutations. Within the context of the 
present paper, artificial DNA of this kind could ease the stringent
constraints imposed on $W$, as was indicated above.

It is worth noting that the "focusing" of solutions (see Fig.13)
may have a very important bearing on mutations.
In fact, it amounts to the
possibility of a weak external influence generating a low amplitude
distribution of mutation sites that would focus itself later
on a high amplitude distribution concentrated in a different region
of the molecule. Thus, one may expect generating mutations
by low intensity agents
distributed in a region of the molecule, or to put it the other way round,
acting on a set of codons different from those that suffer the actual
mutation.

Finally we would like to point out that the irradiation with
electro-magnetic waves at frequencies of the proton tunneling in hydrogen 
bonds  may result in tautomeric transitions of nucleotides, and,
according to the arguments given in this paper, mutations.
This circumstance could be used for experimental verification 
of our model.

The authors are thankful to V. Benderskii and R.Lavery for interesting 
discussion.

\vspace{1cm}

\noindent
{\bf Appendix - Radiation Filtering (RF) Algorithm}

The method used in this paper for constructing soliton solutions is 
based on the following heuristic argument. Suppose we have a solution 
that is close to a soliton, then we may visualize it as a central peak that 
radiates waves of small amplitude (see Fig.10), during the evolution in time. 
In case of soliton proper, the radiation is absent, so that the radiation 
is a specific feature that gives a measure of not being a 
soliton, so that the difference between the central peak and the 
soliton is carried away by the radiation. Hence, the soliton solution 
may be obtained by annihilating the radiation through the 
use of an appropriate filter (see Fig. 14).

The actual algorithm runs as follows:
Take the parameter of cutoff, $\Pi$, which determines the 
level of noise to be absorbed, and the parameter $Err$ for
estimating the precision.
\begin{enumerate}
  \item Let us take a trial set of amplitudes $\tilde{B}_n$ 
        $(n = 0 \dots N-1)$, that is the
        real and the imaginary parts Re $\tilde{B}_n$, Im $\tilde{B}_n$
        and the absolute values $|\tilde{B}_n|$.
  \item Consider the evolution of $\tilde{B}_n$ described by equation
        (\ref{schnl}) for the initial values given by the set $\tilde{B}_n$,
        that is $B_n(t=0) = \tilde{B}_n$ $(n = 0 \dots N-1)$
        for the time interval $\Delta t$, and take the set of amplitudes
        $\hat{B}_n = B_n(t=\Delta t)$.
  \item Consider the set of absolute values $|\hat{B}_n|$, and find the
        index $M$ for the maximum value, $B_M$, of $|\hat{B}_n|$ 
        $(n = 0 \dots N-1)$.
  \item Find the indices $L$ and $R$ such that 
        $|\hat{B}_L|$ and $|\hat{B}_R|$ are 
        both local minima of the set $|\hat{B}_n|$, and the constraints \\
        (i)   $|\hat{B}_L| < \Pi$ and $|\hat{B}_R| < \Pi$, \\
        (ii)  $L < M < R$, \\
        (iii) $L$, $R$ are closest to the index $M$, \\
        are verified.
  \item Consider the new set of amplitudes $B_n(t)$
        $$
          B_n(\Delta t) = \left\{ 
                                  \begin{array}{ll}
                                    \displaystyle{
                                      \frac{\hat{B}_n(\Delta t)}{A}
                                    }, & L \leq n \leq R \\
                                    0, & \mbox{otherwise}
                                  \end{array} 
                          \right.
        $$
        where 
        $$
          A = \left[\sum_{n=L}^R |\hat{B}_n (\Delta t)|^2\right]^{\frac{1}{2}}
        $$
  \item Consider $B_n(\Delta t + T)$ which are the solution to
        equation (\ref{schnl}) for the initial values given by
        $B_n(\Delta t)$ and the time interval of integration equal to $T$.
        Verify that the mean square deviation
        $$
          \sum_n \left[|B_n (\Delta t + T) - B_n(\Delta t)\right]^2,
            \qquad \Delta t \ll T
        $$
        is less than the accuracy level $Err$ accepted. In case it is not met,
        set
        $$
          \tilde{B}_n = B_n(t=0)
        $$
        else exit.
  \item Return to step 2.
\end{enumerate}
\vspace{3mm}

In case of solitons with large support, or long wavelengths, much larger 
than the spacing $a$, we can use the usual approach and look for solitons
of the form (\ref{sts}), that is
$$
  Y(x,t) = e^{i(q \cdot na - \omega t)}\, \psi(na -  vt)
$$
On substituting $B_n(t)$ given above in equation (\ref{schnl}), we
obtain the two standard equations for the imaginary and the real parts of
$B_n(t)$. In the long wavelengths limit, the imaginary one results in the
equation for the velocity of soliton, which in our notation reads:
$$
  v = 2 a \cdot \omega_P \cdot \sin (a \cdot q).
$$
The real part gives a nonlinear functional equation for the amplitude $\psi$,
which can be solved by the familiar Newton method.

It is important that, for long wavelengths, the RF-algorithm
brings about the same results as the Newton method indicated above.

\pagebreak

\pagebreak

 \centerline{{\bf Figure Captions}  \vspace{1cm}}

 \noindent
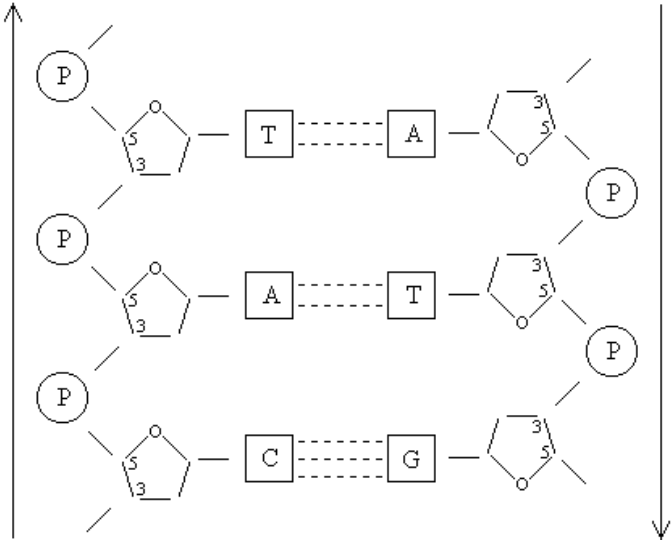
 {\bf Fig.1} \\
 Schematic structure of the double helix of DNA.

 \noindent
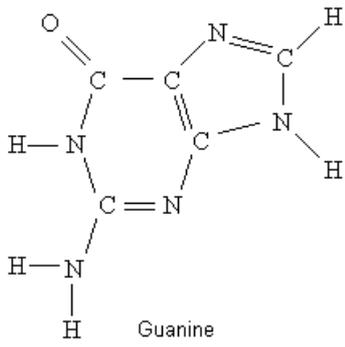
{\bf Fig.2} \\
 Amino/imino and keto/enol forms of purine and pyrimidine.

 \noindent
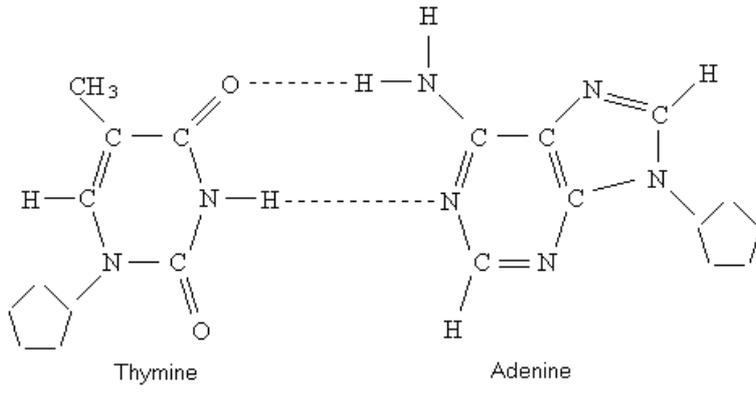
{\bf Fig.3} \\
 Thymine-Adenine and Cytosine-Guanine pairing.

  \noindent
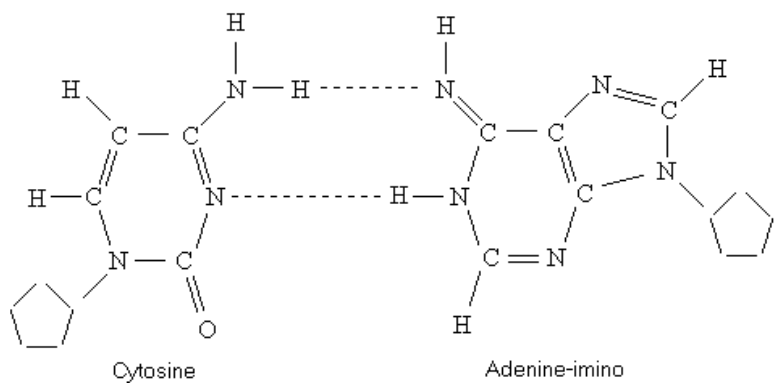
 {\bf Fig.4} \\
 Pairing of cytosine, normal form, and adenine, imino form.

 \noindent
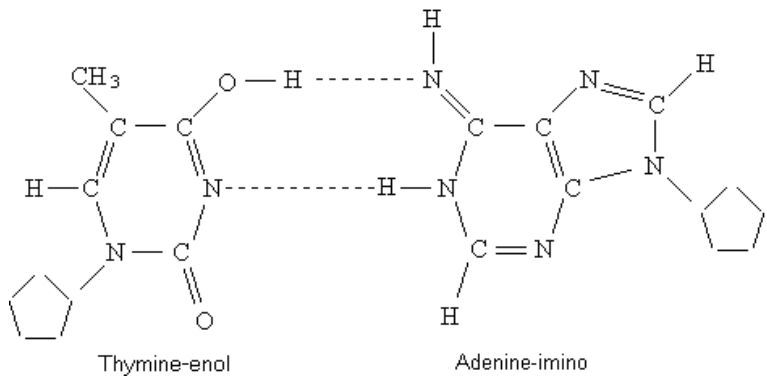
{\bf Fig.5}   \\
 Pairing of thymine, enol form, and adenine, imino form.

 \noindent
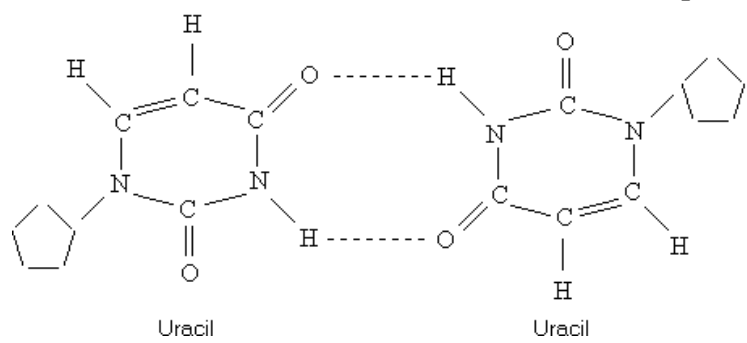
{\bf Fig.6}\\
 Uracil-uracil pairing.

 \noindent
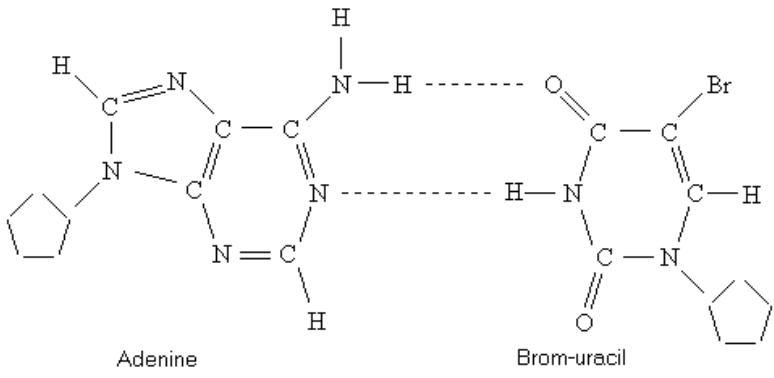
{\bf Fig.7}\\
 Pairing of adenine-(brom-uracil) and guanine-(brom-uracil).

 \noindent
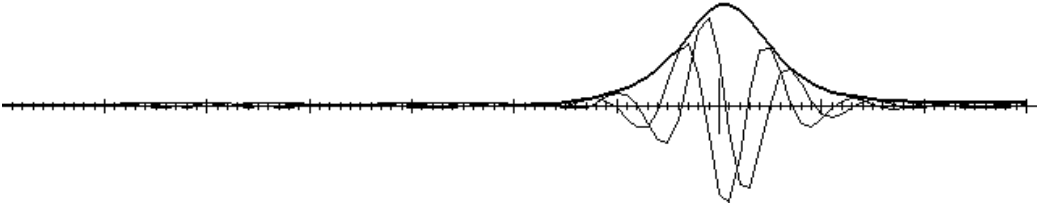
{\bf Fig.8}\\
 Typical moving soliton for $W=0.75, \quad \Lambda= 0.001$,
 velocity $1325.28$ base pairs per $msec$, the period of time spent
 $0.533 \, msec$.  The solid line shows $|A_n|^2$, and the thin lines
 indicate the real and the imaginary part of the amplitude $A_n$.

 \noindent
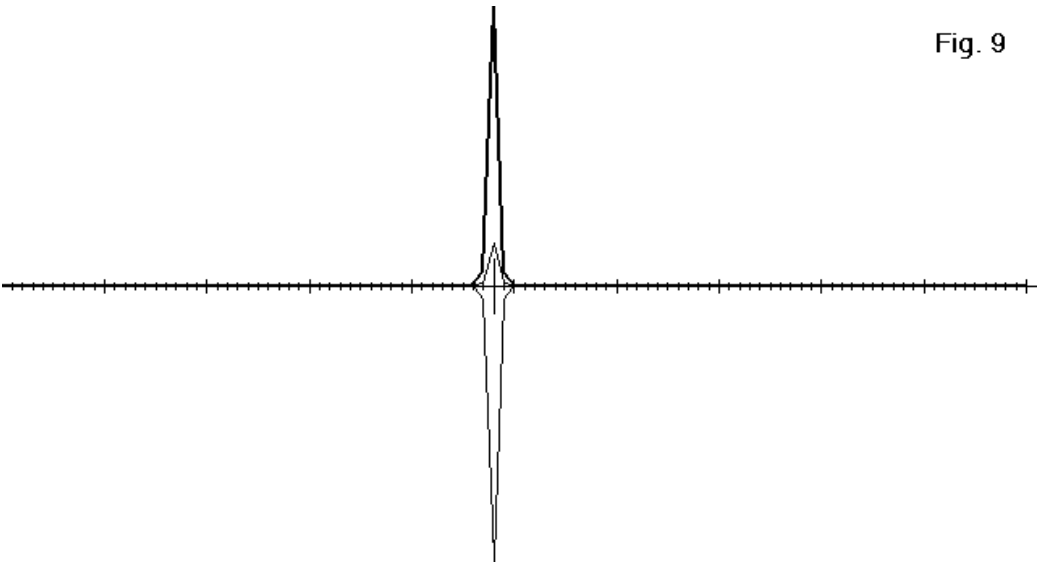
{\bf Fig.9} \\
 Breather, or still soliton.
 Typical moving soliton for $W=10, \quad \Lambda= 0.5$,
 velocity $0$ base pairs per $msec$, the period of time spent $0.501 \, msec$.
 The solid line shows $|A_n|^2$, and the thin lines
 indicate the real and the imaginary part of the amplitude $A_n$.

 \noindent
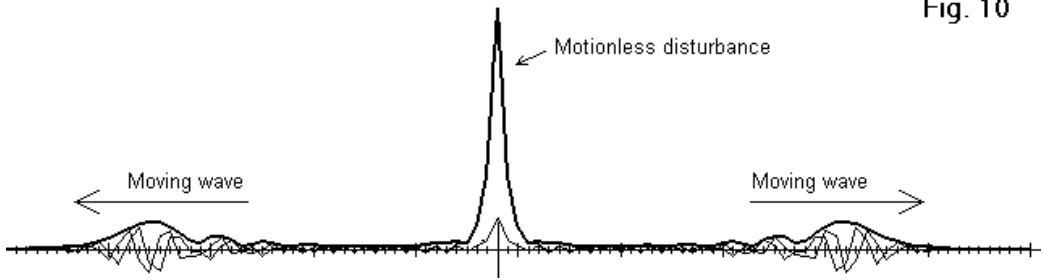
{\bf Fig.10} \\
 Radiation emitted from  the motionless central peak
 during the period of time $0.018$, for $W=5$ and $\Lambda=0.5$;
 the velocity of side waves $1833.3$ base pairs/msec, distance traveled $33$
 base pairs.
 The solid line shows $|A_n|^2$, and the thin lines
 indicate the real and the imaginary part of the amplitude $A_n$.

 \noindent
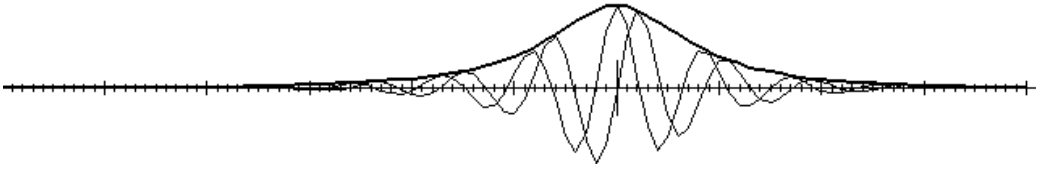
{\bf Fig.11}\\
 Moving soliton for $W=0.75, \quad \Lambda= 0.075$,
 velocity $1335.46$ base pairs per $msec$, the period of time spent
 $0.530 \, msec$, distance travelled $707$ base pairs.
 The solid line shows $|A_n|^2$, and the thin lines
 indicate the real and the imaginary part of the amplitude $A_n$.

 \noindent
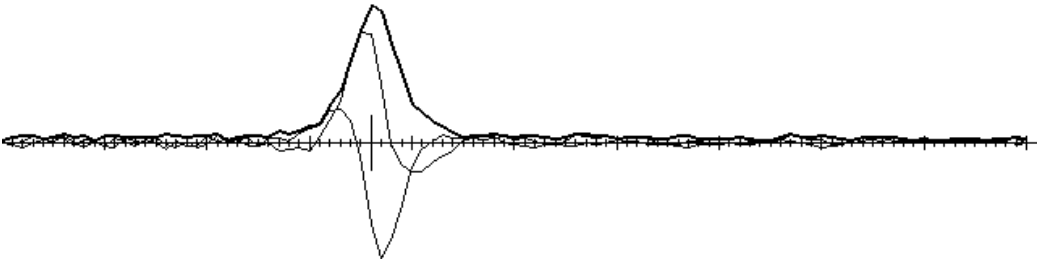
{\bf Fig.12}\\
 Moving soliton for $W=2, \quad \Lambda= 0.1$,
 velocity $851.2$ base pairs per $msec$, the period of time spent
 $0.577 \, msec$, distance travelled $491$ base pairs.
 The solid line shows $|A_n|^2$, and the thin lines
 indicate the real and the imaginary part of the amplitude $A_n$.
 The values of $W, \, \Lambda$ are close to the borderline (see Fig.17),
 dividing the region of stable solitons from the unstable ones.

 \noindent
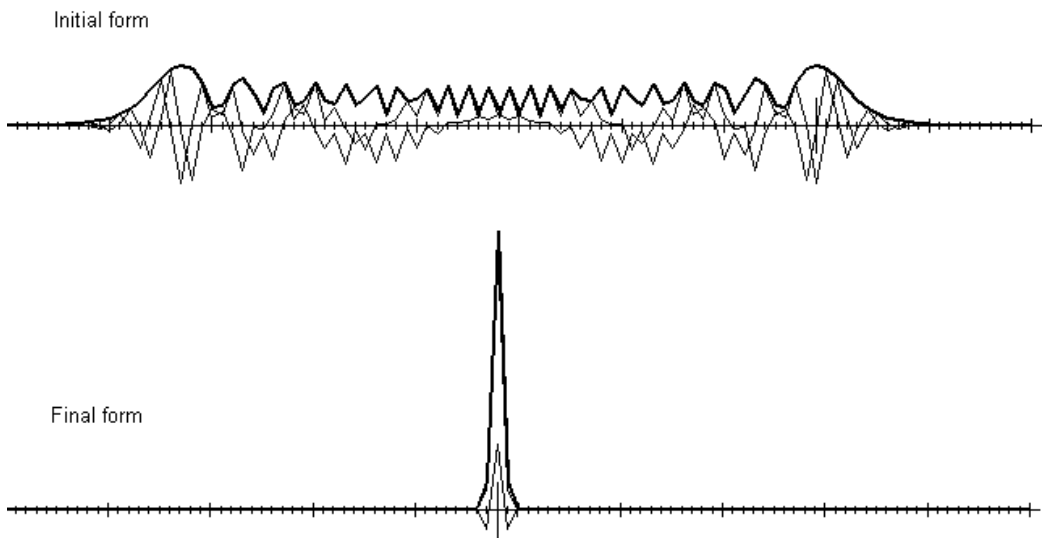
{\bf Fig.13}\\
 Partial self-focusing of an initial low amplitude distribution on a peak
 for a period of time $0.002 \, msec$, for $W=1, \, \Lambda = 0.5$.

 \noindent
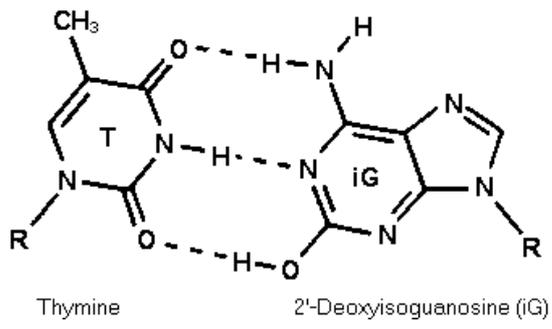
{\bf Fig.14}\\
 Pairing of thymine and $2^{\prime}$-Deoxyisoguanosine.

 \noindent
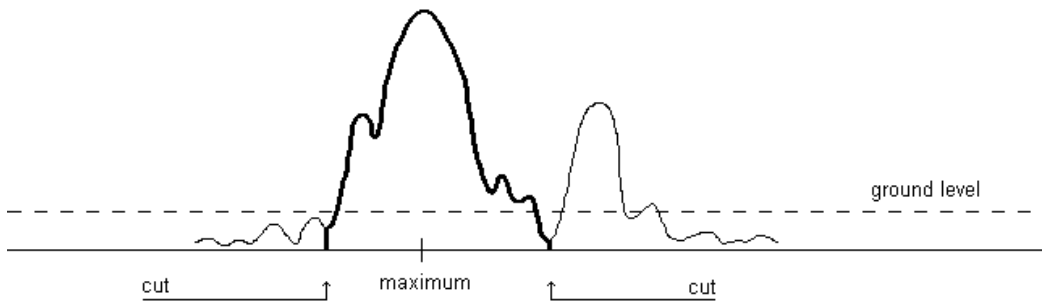
{\bf Fig.15}\\
 Filter algorithm for finding solitons.
 The solid line indicates the part of the excitation to be preserved,
 and the thin line the does the part to be cut off,
 the dashed line means the precision.

 \noindent
{\bf Fig.16 (a)-(d)}\\
 Sets of $W$ and $\nu$, which allow for soliton solutions, at fixed 
 $\Lambda$: \\
 (a) for $\Lambda = 0$ \\
 (b) for $\Lambda = 0.1$ \\
 (c) for $\Lambda = 0.15$ \\
 (d) for $\Lambda = 0.2$ \\
 The solid line shows still solitons, or breathers, and the shaded area
 indicates moving solitons. We take into account only solitons of size
 less than 100 base pairs.
 
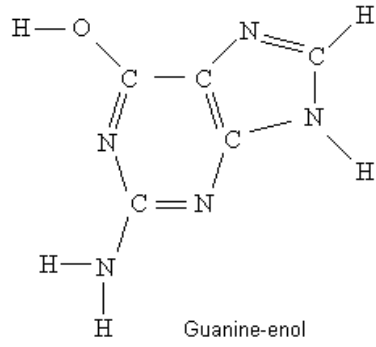
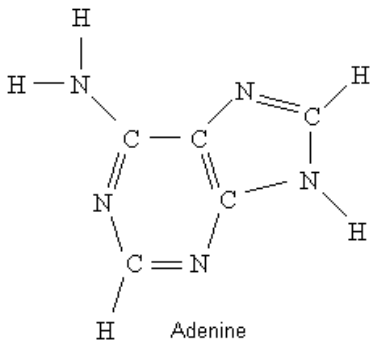
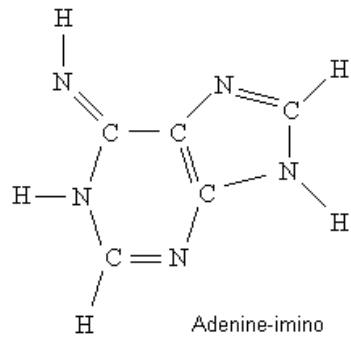
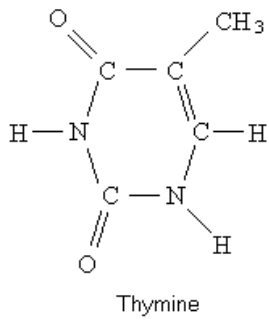
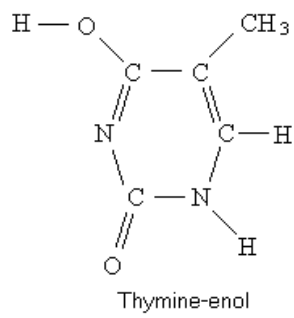
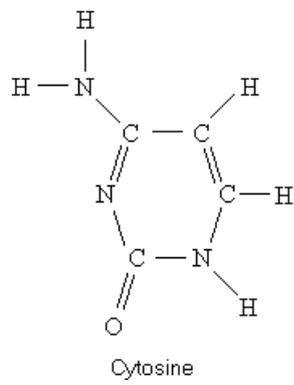
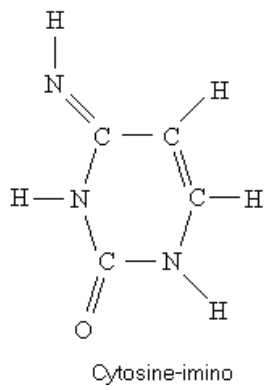
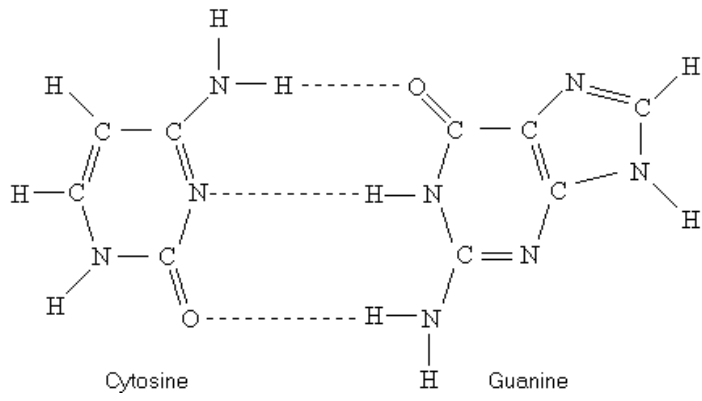
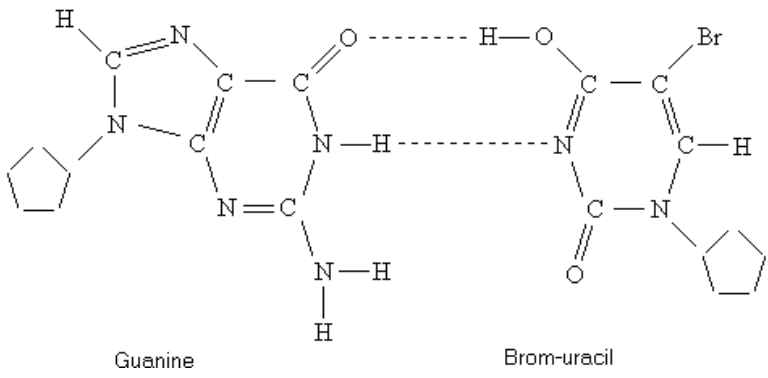
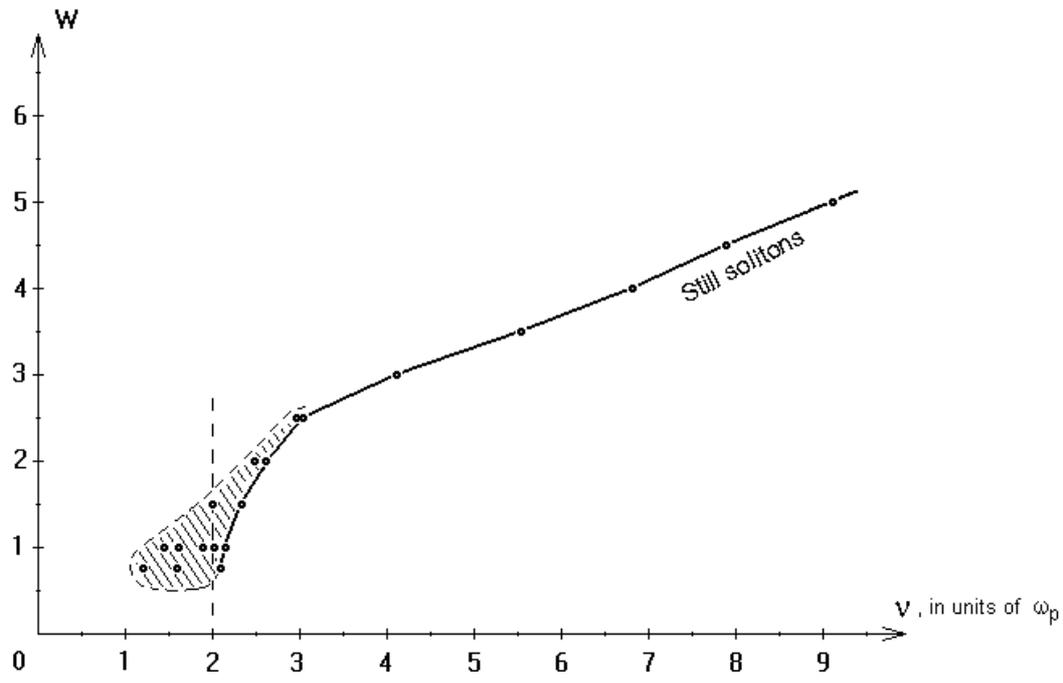
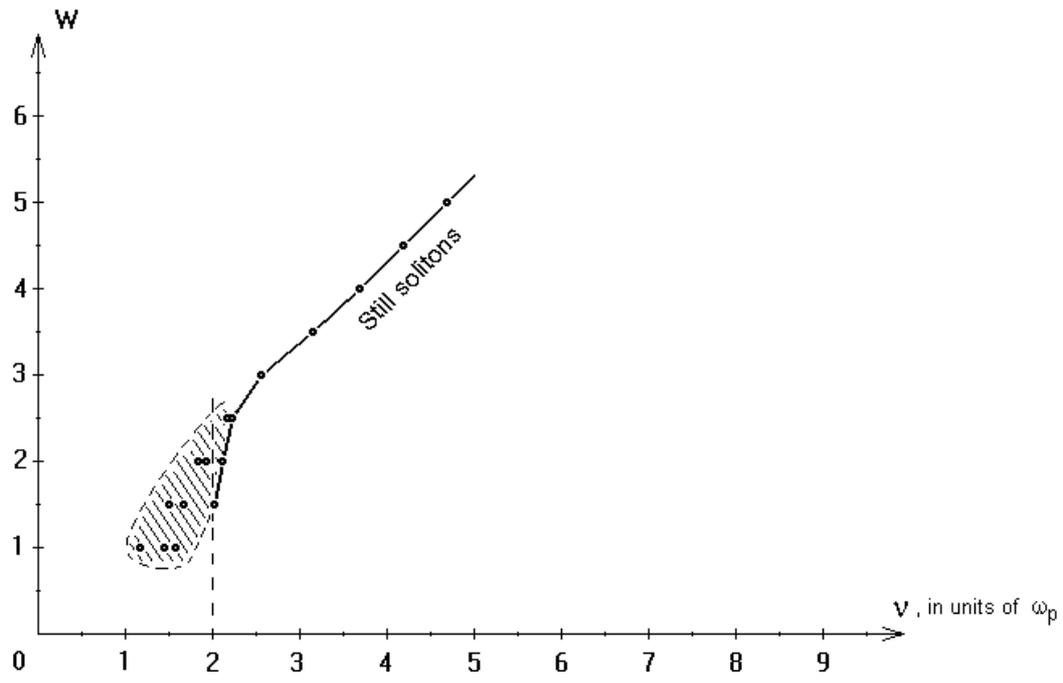
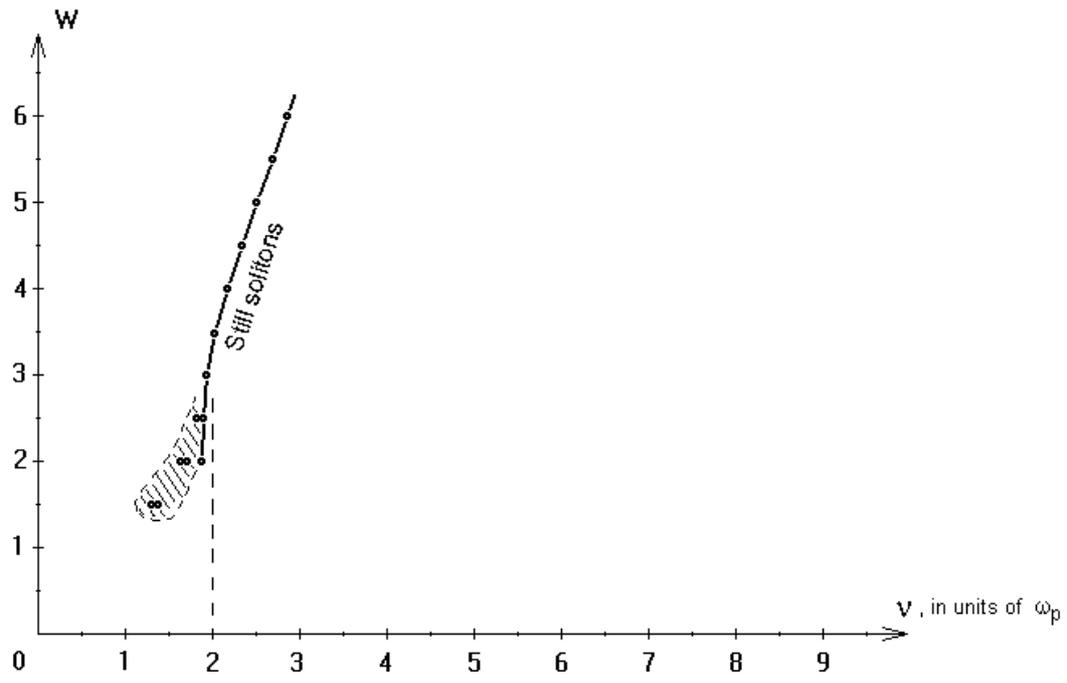
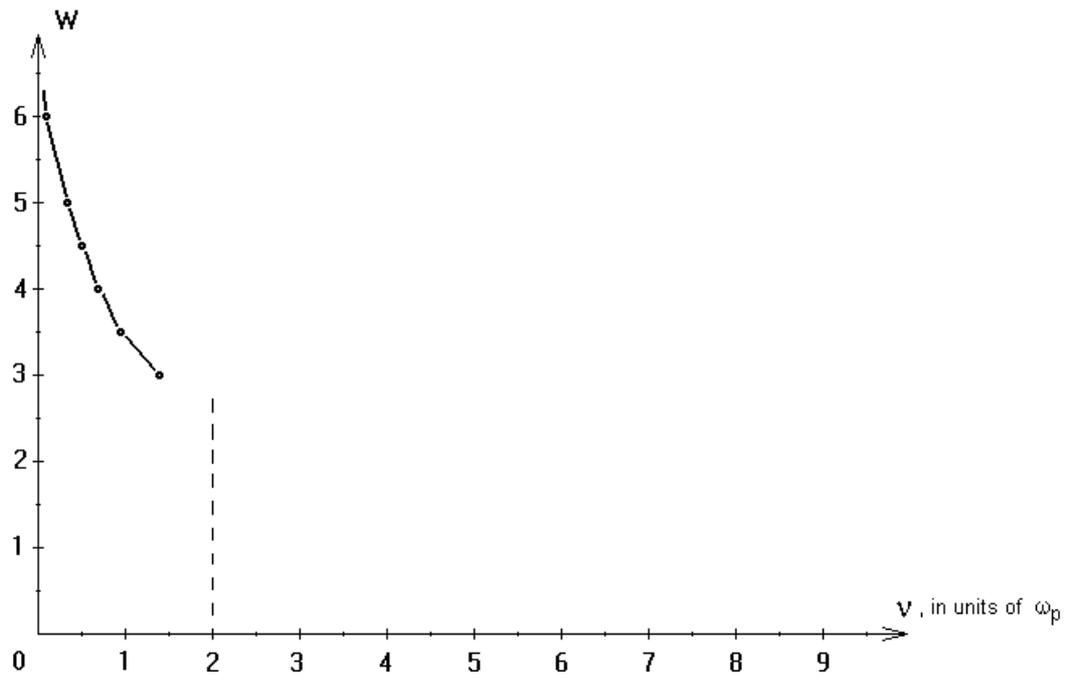
\end{document}